\documentclass[aps,twocolumn,superscriptaddress,showkeys,secnumarabic]{revtex4}
\usepackage[pdftex]{graphicx}

\pdfoutput=1

\usepackage[pdftex]{hyperref}

\begin{document}

\title{Chaotic singular maps}
\author{M. G. Cosenza}
\affiliation{Centro de F\'isica Fundamental, Universidad de Los Andes, M\'erida, Venezuela}
\author{O. Alvarez-LLamoza}
 \affiliation{Departamento de F\'{\i}sica, FACYT, Universidad de
        Carabobo, Valencia, Venezuela}
\affiliation{Centro de F\'isica Fundamental, Universidad de Los Andes, M\'erida, Venezuela}

\begin{abstract}
We consider a family of singular maps as an example of a simple model of dynamical systems exhibiting the property of robust chaos on a well defined range of parameters.  Critical boundaries separating the region of robust chaos from the region where stable fixed points exist are calculated on the parameter space of the system. It is shown that the transitions to robust chaos in these systems occur either through the routes of type-I or type-III intermittency and the critical boundaries for each type of transition have been determined on the phase diagram of the system. The simplicity of these singular maps and the robustness of their chaotic dynamics make them useful ingredients in the construction of models and in applications that require reliable operation under chaos.
\end{abstract}


\maketitle

\section{Introduction}

Many practical uses of the phenomenon of chaos have been proposed in recent years, as for instance, in communications \cite{hayes,argyris}, in enhancing mixing in chemical processes \cite{ottino}, in avoiding electromagnetic interferences \cite{deane}, in cryptography \cite{baptista}, in stabilizing plasma fusion \cite{chandre}, etc. In such applications it is necessary to obtain reliable operation of chaotic systems.

It is known that most chaotic attractors of smooth systems are embedded with a dense set of periodic windows for any range of parameter values. Therefore in practical systems functioning in chaotic mode, a slight fluctuation of a parameter may drive the system out of chaos. On the other hand, it has been shown that some dynamical systems can exhibit robust chaos \cite{banerjee,potapov,priel}.  A chaotic attractor is said to be robust if, for its parameter values, there exist a neighborhood in the parameter space with absence of periodic windows and the chaotic attractor is unique \cite{banerjee}. Robustness is an important property in applications that require reliable operation under chaos in the sense that the chaotic behavior cannot be destroyed by arbitrarily small perturbations of the system parameters. For example, robust chaos has efficiently been used in communications schemes \cite{garcia}.

In this article we study a family of singular maps as an example of a simple model of dynamical systems that shows robust chaos on a finite interval of their parameter values. In Section 2 we introduce this family of maps and investigate their dynamical properties, both analytically and numerically. It is found that the transitions to robust chaos in these systems occur either through the routes of type-I or type-III intermittency \cite{pomeau}. The region where robust chaos takes place is characterized on the space of parameters of the maps. Conclusions are presented in Section 3.

\section{Singular maps}

As a simple model of a dynamical system displaying robust chaos, we consider the following family of singular maps
\begin{equation}
x_{n+1}=f(x_n)=b-|x_n|^z ,
\label{sing_map}
\end{equation}
where $n \in Z$, $|z| < 1$, and $b$ is a real parameter. The exponent $z$ describes the order of the singularity at the origin that separates two piecewise smooth branches of the map Eq. (\ref{sing_map}).  These maps are unbounded, that is, $x_n \in (-\infty,\infty)$. The Schwarzian derivative of the family of maps Eq. (\ref{sing_map}) is always positive, i.e.,
\begin{equation}
Sf=\frac{f'''}{f'}-\frac{3}{2}\left( \frac{f''}{f'}\right) ^2=\frac{1-z^2}{2x^2}>0.
\end{equation}
for $|z| < 1$. Thus maps defined by Eq. (\ref{sing_map}) do not belong to the standard universality classes of unimodal maps and do not satisfy Singer’s theorem \cite{singer}. As a consequence, these singular maps do not exhibit a sequence of period-doubling bifurcations. Instead, the condition $Sf>0$ leads to the occurrence of an inverse period-doubling bifurcation, where a stable fixed point on one branch of the singular map losses its stability at some critical value of the parameter $b$ to yield robust chaos. It should be noted that robust chaos has also been discovered in smooth, continuous one-dimensional maps \cite{andrecut}.

\begin{figure}[t]
\includegraphics[scale=0.8]{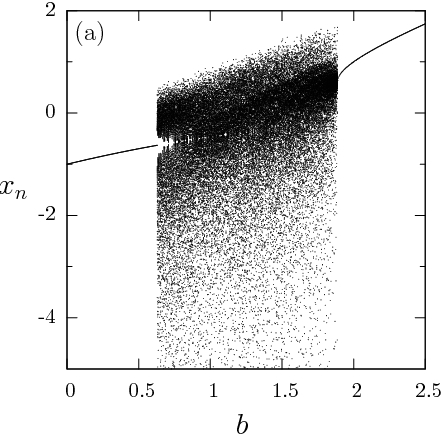}\\
\vspace{0.4cm}
\includegraphics[scale=0.8]{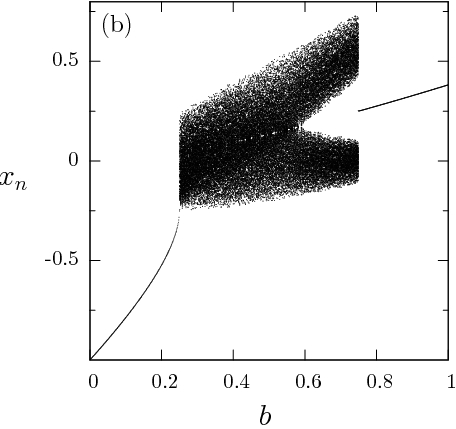}
\caption{Bifurcation diagrams of the iterates of the map Eq. (\ref{sing_map})  as a function of the parameter $b$ for two values of the order of the singularity $z$, showing robust chaos. Type-I or type-III intermittencies appear at the boundaries of the robust chaos intervals. (a) $z = -0.5$;  (b) $z = 0.5$.}
\label{bif_diag}
\end{figure}

Figure \ref{bif_diag} shows the bifurcation diagrams of the iterates of map Eq. (\ref{sing_map}) as a function of  the parameter $b$ for two different values of the singularity exponent $z$. Figure \ref{bif_diag} reveals robust chaos, i.e., the absence of windows of stable periodic orbits and coexisting attractors, on a well defined interval of the parameter $b$ for each value of  $z$.

The transition to chaos at the boundaries of the robust chaotic interval occurs by intermittency. Intermittent chaos is characterized by the display of long sequences of periodiclike behavior, called the laminar phases, interrupted by comparatively short chaotic bursts. The phenomenon has been extensively studied since the original work of Pomeau and Manneville \cite{pomeau} classifying type-I, -II, and -III instabilities when the Floquet multipliers of the local Poincar\'e map associated to the system crosses the unit circle. Type-I intermittency occurs by a tangent bifurcation when the Floquet’s multiplier for the Poincar\'e map crosses the circle of unitary norm in the complex plane through $+1$; type-II intermittency is due to a Hopf’s bifurcation which appears as two complex eigenvalues of the Floquet’s matrix cross the unitary circle off the real axis; and type-III intermittency is associated to an inverse period doubling bifurcation whose Floquet’s multiplier is $-1$.

\begin{figure}[t]
{\centerline{\includegraphics[scale=0.8]{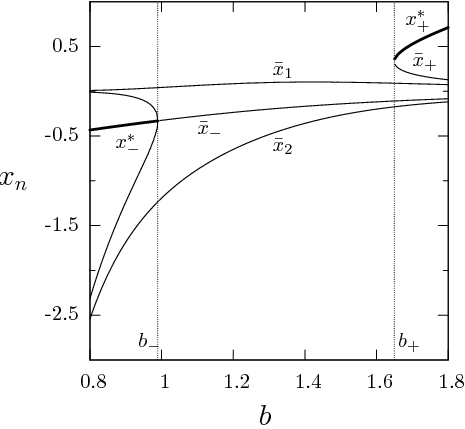}}}
\caption{Some unstable periodic orbits of the singular map with $z = -0.25$, indicated by dotted lines, as a function of $b$. The stable fixed points $x^*_{-}$ and $x^*_{+}$ are plotted with solid lines. At the parameter value $b_{-}=0.9896$, the fixed point $x^*_{-}$ becomes unstable through an inverse period-doubling bifurcation, giving raise to the unstable fixed point $\overline{x}_{-}$. At the value $b_{+}=1.6494$, a tangent bifurcation takes place and the pair of points $x^*_{+}$ (stable) and $\overline{x}_{+}$ (unstable) are born. The period-2 unstable orbit $\overline{x}_{1}$ and $\overline{x}_{2}$, satisfying $f(\overline{x}_{1})=f(\overline{x}_{2})$, are shown.}
\label{unst_orbits}
\end{figure}

Two stable fixed points satisfying $f(x^*)=x^*$ and $|f'(x^*)|<1$ exist for each value of $z$: $x^*_{-}<0$ and $x^*_{+}>0$, both are seen in Figure \ref{bif_diag}. For $z \in (-1,0)$, the fixed point  $x^*_{-}$ becomes unstable at the parameter value
\begin{equation}
b_{-}(z)=|z|^{\frac{z}{1-z}}-|z|^{\frac{1}{1-z}},
\end{equation} 
through an inverse period doubling bifurcation that gives rise to chaos via type-III intermittency,  while the fixed point $x^*_{+}$ originates from a tangent bifurcation at the value
\begin{equation}
b_{+}(z)=|z|^{\frac{z}{1-z}}+|z|^{\frac{1}{1-z}},
\end{equation} 
and the transition to chaos at this value of $b$ takes place through type-I intermittency. On the other hand, for $z \in (0,1)$ the behavior of the fixed points is interchanged: $x^*_{-}$ experiences a tangent bifurcation at the parameter value $b_{-}(z)$ and a type-I intermittent transition to chaos occurs; while the fixed point $x^*_{+}$ undergoes an inverse period-doubling bifurcation at the value $b_+(z)$, setting the scenario for a type-III intermittent transition to chaos. There exist several unstable period-$m$ orbits $\{ \overline{x}_1,\overline{x}_2,\ldots ,\overline{x}_m\}$ satisfying $f^{(m)}(\overline{x}_j)=\overline{x}_j$ and $|\frac{d}{dx}f^{(m)}(\overline{x}_j)|=\prod^m_{j=1}|f'(\overline{x}_j)|<1$ in the chaotic interval $b \in [b_{+}(z),b_{-}(z)]$. Figure \ref{unst_orbits} shows some unstable periodic orbits of the singular map with $z = -0.25$ as a function of the parameter $b$.

\begin{figure}[t]
{\centerline{\includegraphics[scale=0.8]{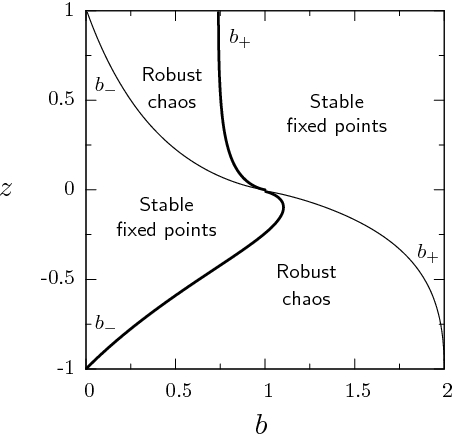}}}
\caption{Critical boundaries $b_-(z)$ and $b_+(z)$ of the robust chaos region for the singular maps on the space of parameters $(b,z)$. The thick, dark line indicates the boundary $b_{III}(z)$ for the transition to chaos via type-III intermittency. The thin, light line corresponds to the boundary $b_{I}(z)$ for the onset of type-I intermittency.}
\label{crit_bound}
\end{figure}

Figure \ref{crit_bound} shows the critical boundaries $b_{-}(z)$ and $b_{+}(z)$ for the transition to chaos. These boundaries separate the region on the parameter plane $(b,z)$ where robust chaos takes place from the region where stable fixed points of the maps Eq. (\ref{sing_map}) exist. The transition to chaos via type-I intermittency takes place at the parameter boundaries $b_I(z)=b_+(z)$ for $z \in (-1,0)$, and $b_I(z)=b_-(z)$ for $z \in (0,1)$. On the other hand, the transition to chaos via type-III intermittency occurs at the critical parameter values $b_{III}(z)=b_-(z)$ for $z \in (-1,0)$, and $b_{III}(z)=b_+(z)$ for $z \in (0,1)$. The boundaries $b_I(z)$ and $b_{III}(z)$ on the space of parameters $(b,z)$ are indicated in Figure \ref{crit_bound}. The width of the interval for robust chaos on the parameter $b$ for a given $|z|<1$ is
\begin{equation}
 \Delta b(z)=b_+(z)-b_-(z)=2|z|^{\frac{1}{1-z}}.
\end{equation}

\begin{figure}[th]
\includegraphics[scale=0.8]{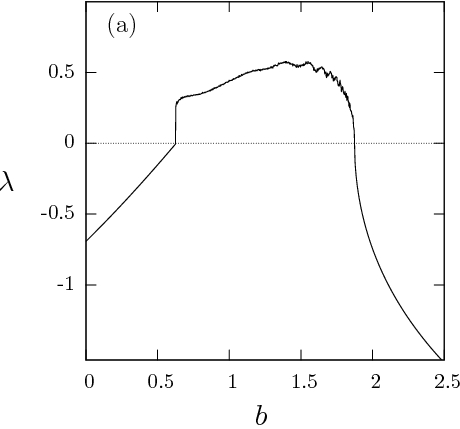}\\
\vspace{0.4cm}
\includegraphics[scale=0.8]{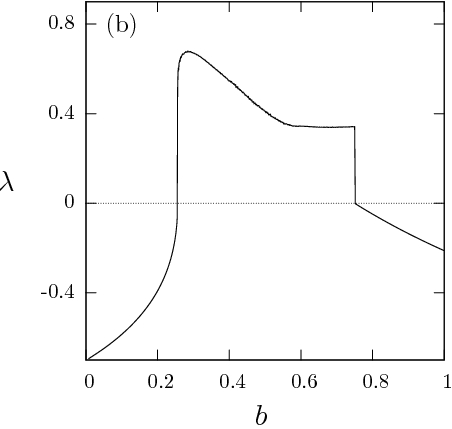}
\caption{Lyapunov exponent $\lambda$ as a function of the parameter $b$ for two values of $z$, calculated over $5 \times 10^4$ iterations after neglecting $5 \times 10^3$ iterates representing transient behavior for each value of $b$. (a) $z = -0.5$; (b) $z = 0.5$.}
\label{lyap_exp}
\end{figure}

Figure \ref{lyap_exp} shows the Lyapunov exponent $\lambda$ as a function of the parameter $b$ for the family of maps Eq. (\ref{sing_map}), for two values of $z$, calculated as
\begin{equation}
\lambda = \frac 1 T \sum_{n=1}^{T} \log \left| f'(x_n) \right|,
\label{exp_lyap}
\end{equation}
with $T=5 \times 10^4$ iterates after discarding $5000$ transients for each parameter value. The boundaries $b_-(z)$ and $b_+(z)$ correspond to the values $\lambda=0$. The Lyapunov exponent is positive on the robust chaos interval $\Delta b(z)$. The transition to chaos through type-I intermittency is smooth, as seen in Figure \ref{lyap_exp}. In contrast, the transition to chaos via type III intermittency is manifested by a discontinuity of the derivative of the Lyapunov exponent at the parameter values corresponding to the critical boundary $b_{III}(z)$. This discontinuity is due to the sudden loss of stability of the fixed point associated to the inverse period doubling bifurcation that occurs at the boundary $b_{III}(z)$. The Lyapunov exponent can be regarded as an order parameter that characterizes the transition to chaos via type-I or type-III intermittency.  This transition can be very abrupt in the case of type-III intermittency, as seen in Figure \ref{lyap_exp}.

\section{Conclusions}

We have introduced a family of singular maps as an example of a simple model of dynamical systems exhibiting robust chaos on a well defined range of parameters.  The behavior of these maps has been characterized as a phase diagram in the space of their parameters, showing a region where robust chaos takes place and regions where stable fixed points occur. We have shown that the transitions to robust chaos in these systems occur either through the routes of type-I or type-III intermittency and have calculated the critical boundaries for each type of transition on the phase diagram of the systems. The simplicity of these singular maps and the robustness of their chaotic dynamics make them useful ingredients in the construction of models and in applications that require the property of chaos.

\begin{acknowledgments}
This work was supported by Consejo de Desarrollo Cient\'ifico, Human\'istico y Tecnol\'ogico of  the Universidad de Los Andes, M\'erida,  under grant No. C-1396-06-05-B and by FONACIT, Venezuela, under grant No. F-2002000426.
\end{acknowledgments}

\end{document}